\documentstyle[twoside,epsfig]{pro12}
\begin{document}
%
\typeout{============================================================================}
\typeout{ WRITING: Contribution of H. Onel (onel@physik.tu-berlin.de - honel@aip.de) }
\typeout{============================================================================}
%
%
\author{H.~\"O{}nel\adress{\sl Astrophysikalisches Institut Potsdam (AIP), An der Sternwarte 16, D--14482 Potsdam, Germany}$\;\,^{\dagger}\,$, 
        G.~Mann$^*\,$ and E.~Sedlmayr\adress{\sl  Technische Universit\"a{}t Berlin (TU--Berlin), Zentrum f\"u{}r Astronomie und Astrophysik, Sekr. PN 8-1, Hardenbergstra\ss{}e 36, D--10623 Berlin, Germany}}
\title{TRANSPORT OF ENERGETIC ELECTRONS THROUGH THE SOLAR CORONA AND THE INTERPLANETARY SPACE}
\head{H.~\"O{}nel, G.~Mann, and E.~Sedlmayr}{Transport of Energetic Electrons Through the Solar Corona \& the Interplanetary Space}
\date{22.~April~2005}
%
%
\maketitle
\newcommand{\dd}[1]{{\rm d}#1}
\newcommand{\ddfrac}[2]{\frac{{\rm d}#1}{{\rm d}#2}}
\newcommand{\dt}[1]{\frac{{\rm{d}}#1}{{\rm d}t}} 
\newcommand{\ddt}[1]{\frac{{\rm d}^2#1}{{\rm d}t^2}} 
\newcommand{\gefordert}{\stackrel{!}{=}} 
\newcommand{\mysun}[0]{{\rm Sun}}
\newcommand{\Rs}[0]{R_\mysun}
\def\ga{\mathrel{\mathchoice {\vcenter{\offinterlineskip\halign{\hfil 
         $\displaystyle##$\hfil\cr>\cr\sim\cr}}}
         {\vcenter{\offinterlineskip\halign{\hfil$\textstyle##$\hfil\cr
         >\cr\sim\cr}}}
         {\vcenter{\offinterlineskip\halign{\hfil$\scriptstyle##$\hfil\cr
         >\cr\sim\cr}}}
         {\vcenter{\offinterlineskip\halign{\hfil$\scriptscriptstyle##$\hfil\cr
         >\cr\sim\cr}}}}}
\def\la{\mathrel{\mathchoice {\vcenter{\offinterlineskip\halign{\hfil 
        $\displaystyle##$\hfil\cr<\cr\sim\cr}}}
        {\vcenter{\offinterlineskip\halign{\hfil$\textstyle##$\hfil\cr  
        <\cr\sim\cr}}}
        {\vcenter{\offinterlineskip\halign{\hfil$\scriptstyle##$\hfil\cr
        <\cr\sim\cr}}}
        {\vcenter{\offinterlineskip\halign{\hfil$\scriptscriptstyle##$\hfil\cr
        <\cr\sim\cr}}}}}
\begin{abstract}
During solar flares fast electron beams generated in the solar corona
are \mbox{non-thermal} radio sources in terms of type~III bursts. Sometimes 
they can enter into the interplanetary space, where they can be observed 
by in-situ measurements as it is done \mbox{e.g.} by the \textit{WIND} spacecraft. 
On the other hand, they can be the source of non-thermal X-ray 
radiation as \mbox{e.g.} observed by \textit{RHESSI}, if they precipitate toward 
the dense chromosphere due to bremsstrahlung.
Since these energetic electrons are generated in the corona and observed
at another site, the study of transport of such electrons in the corona
and interplanetary space is of special interest. The transport of
electrons is influenced by the global magnetic and electric field
as well as local Coulomb collisions with the particles in the background
plasma.
\end{abstract}
\section{Introduction}\label{sec_introduction}
\begin{figure}[!ht]
\begin{minipage}[h]{0.55\textwidth}
Both methods provide the energies of the electrons
either at the X-ray emission site (which is different from
the electron acceleration site) or at the current position 
of the spacecraft in the interplanetary medium.
With this information and the model explained in the 
present paper, it is possible to calculate the altitude
of the acceleration site and the kinetic energy that the electrons
would have had there. 
\caption[Electron Propagation]{{Left:} The dynamic spectrum shows type~III (and U) radio bursts in the solar corona, which were detected by the AIP spectral polarimeters. (The intensity is colour 
                                 coded.)
                               {Right:} The propagation of electrons is schematically illustrated. The fast propagating electrons excite type~III radio emissions.}
\end{minipage}
\hfill
\begin{minipage}[h]{0.4\textwidth}
   \begin{center}
    \hspace*{-0.10\textwidth}\includegraphics[width=1.2\textwidth]{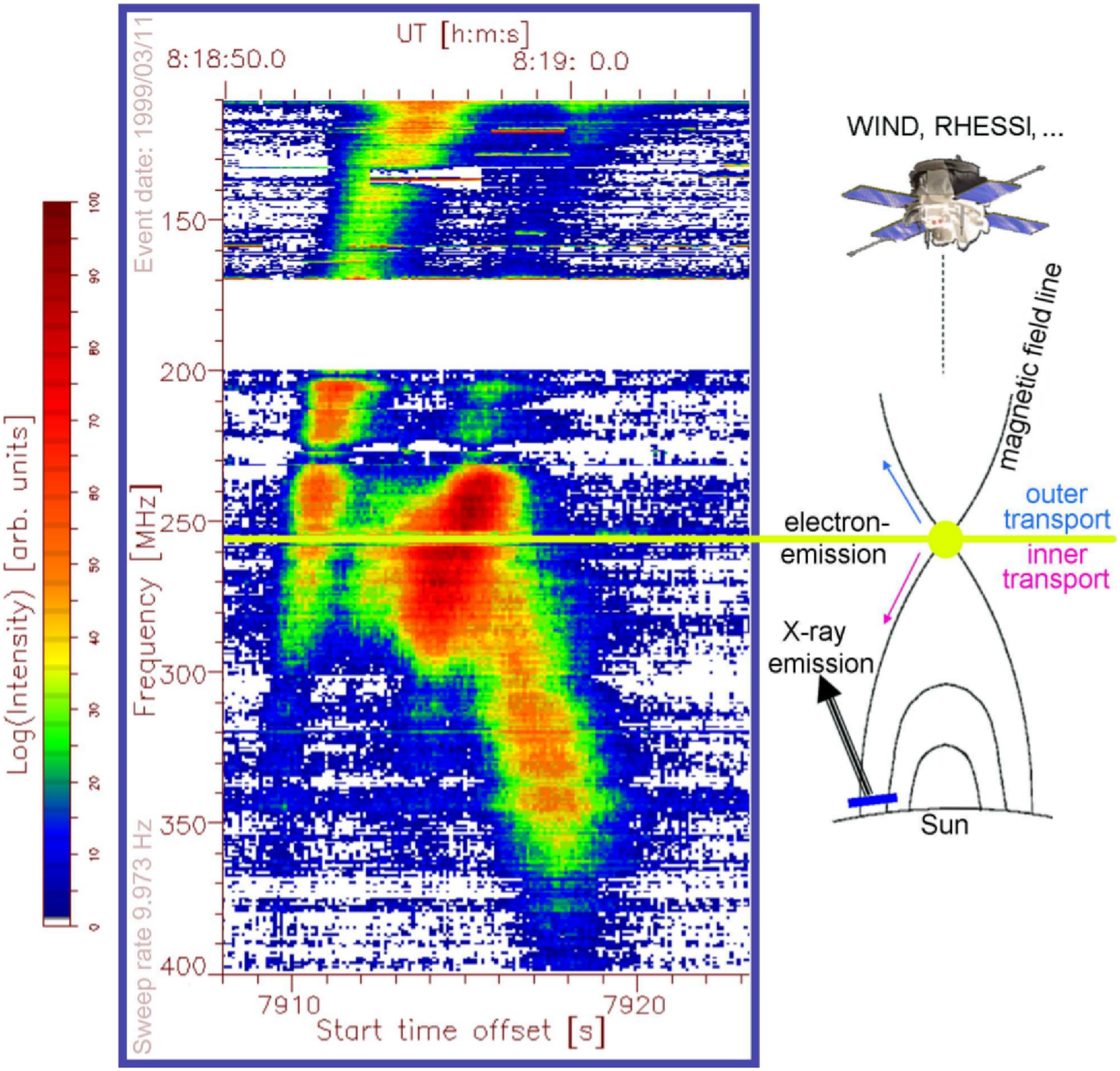} 
    \label{img_aim}
   \end{center}
   \vspace{-1.5cm}\hspace{0.73\textwidth}{\tiny \mbox{Reference: \"O{}nel~(2004)}}\hspace{-0.73\textwidth}\vspace{1.5cm}
\end{minipage}
\end{figure}

In the solar corona energetic electrons are released 
\mbox{e.g.} during solar flares
and travel along magnetic field lines 
either toward or away from the Sun (Figure~\ref{img_aim}).
While they propagate through the plasma background
they excite Langmuir waves via \mbox{beam-plasma} 
instability (see \mbox{e.g.} Melrose,~(1985)). Partly the energy of 
Langmuir waves is converted into electromagnetic radiation 
with a frequency close to the electron plasma frequency $f$
\begin{eqnarray}
f & = & \sqrt{\frac{e^2 N_{{\rm e}}}{4 \pi^2 \epsilon_0 m_{\rm e}}} \label{eqn_electronplasmafrequency}    
\end{eqnarray}
and/or its harmonics, 
where $\epsilon_0$, $e$, $m_{\rm e}$ and $N_{\rm e}$ denote 
the electric constant, the elementary charge, the mass of an electron 
and the electron number density, respectively.

If those electrons have sufficient energy, and the 
plasma background is dense enough (as in
the case of the solar chromosphere), they can 
emit non-thermal X-ray radiation via bremsstrahlung (see \mbox{e.g.} Brown,~(1972)), 
which can be detected by spacecrafts as {\sl RHESSI}.
On the other hand electrons moving toward the interplanetary medium up to \mbox{$1~{\rm AU}$}
can be observed in-situ by spacecrafts as {\sl WIND}.
\newpage
In Section~\ref{sec_model} the theoretical model to describe 
the electron propagation through a given plasma 
background will be introduced. Next, the plasma background itself 
is going to be explained in Section~\ref{sec_background}. Finally
in Section~\ref{sec_discussion} some numerical 
solutions will be presented and in Section~\ref{sec_summary} 
the main issues summarised.
\section{Model for Electron Transport}\label{sec_model}
The equations of motion for an electron 
propagating through a given plasma background 
(the latter explained in Section~\ref{sec_background})
will be briefly introduced within the current Section, 
by taking Coulomb collisions, global magnetic and electric
fields into account. 

Therefore, first the assumptions are explained, which are made to 
obtain the equations of motion and next the important 
relations and appearing quantities.

\subsection*{Assumptions} 
\begin{enumerate}
 \item Because collisions are easily to treat in the {\it centre of mass system},
       all calculations are done in this frame of reference.
       This means, all quantities describe only an effective 
       one-particle system consisting of the electron and the 
       plasma background particles.
 \item Within this paper a spherical, heliocentric coordinate system is used consistently.
 \item The electric and the nonuniform magnetic fields are 
       treated to be {\it collinear} and {\it one dimensional}. 
       They are introduced within Section~\ref{sec_background} 
       and have only {\it small spatial variations}, 
       \mbox{i.e.} the {\it magnetic moment becomes an 
       adiabatic constant of motion}. \label{assumption_collinear}
 \item The acceleration by Coulomb collisions is considered
       to be caused under a {\it small angle scattering approximation},
       due to the fact, that large angle scattering phenomena are
       rare in the cases presented within this paper.
\end{enumerate}

\subsection*{Important Relations, Quantities and the Final Equations of Motion} 
Due to Coulomb collisions, the mean squared scattering 
angle \mbox{$\overline{\theta^2}_j$}
will change, according to Jackson~(1962), with the time $t$ as
\begin{eqnarray}
 \dd{\overline{\theta^2}_j} & \approx & \left| \frac{e^2 q_j^2 N_j} {2\pi\epsilon_0^2 \mu_j^2 v_{j,\parallel}^3} \ln\left[\sqrt{\frac{\lambda_{\rm D}^2+b_{0,\parallel}^2}{2 b_{0,\parallel}^2}}\right] \right| \dd{t},\label{eqn_squaredmeandeflectionangle}
\end{eqnarray}
if the electron during its passage through the
background plasma is affected by a particle $j$.
Here $j$ represents a particle of the plasma background, 
with charge \mbox{$q_j$} and particle number density $N_j$, 
while \mbox{$v_j$} and \mbox{$\mu_j$} 
stand for the relative velocity and the effective mass between the 
background particle and the propagating electron. The other appearing
quantities are the Debye length \mbox{$\lambda_{\rm D}=(\epsilon_0 k_{\rm{}B} T (N_{\rm{}e} e^2)^{-1})^{0.5}$},
the Boltzmann constant \mbox{$k_{\rm{}B}$} and the temperature \mbox{$T$}.
\mbox{$b_0=e \left|q_j\right| (4\pi\epsilon_0\mu_j v_j^2)^{-1}$} characterises
an impact parameter for the effective one particle system, which leads to 
a deflection of \mbox{$0.5~\pi$} and allows to 
define \mbox{$b_{0,\parallel}:=b_{0} \cos^{-2}\left[\Theta_j\right]$} by using
the pitch-angle \mbox{$\Theta_j=\angle\left[\vec{B},{\vec{v}_j}\right]$}.
Hence the following \mbox{root-mean-square} (RMS) quantity is introduced by definition
\begin{eqnarray}
 \dd\theta_{{\rm{C}},j} &:=& \sqrt{\dd\overline{\theta^2}_j}.\label{eqn_rms}
\end{eqnarray}
The effects caused by the electric field $E$
and the magnetic flux density $B$ to the pitch angle $\Theta_j$
are given, as Bai~(1982) suggested, by
\begin{eqnarray}
 \dt{\Theta_j} & = & \frac{\sin[\Theta_j]}{v_j}\frac{e E}{\mu_j}+\frac{v_j\sin[\Theta_j]}{2 B}\ddfrac{B}{r}. \label{eqn_pitchangleevolutionconsideringebfields}
\end{eqnarray}
Because of assumption~\ref{assumption_collinear} stated on page~\pageref{assumption_collinear}
it is fully sufficient to investigate the scalar quantities only.

If one considers more than one particle species $j$
in the plasma background,
as it is described in Section~\ref{sec_background}, 
then one has to introduce the following quantities
\begin{eqnarray}
 \dd{}\theta_{{\rm{C}},{\rm{total}}} & = & \dd{}\theta_{{\rm{C}},{\rm{}e}}+\dd{}\theta_{{\rm{C}},{\rm{}p}}+\dd{}\theta_{{\rm{C}},{{\rm{He}}}^{2+}}\label{eqn_rms_total}\\
 \Theta_{\rm{total}}               & = & \Theta_{{\rm{}e}} +\Theta_{{\rm{}p}}+\Theta_{{\rm{He}}^{2+}} \label{eqn_pitch_total}\\
 \mu_{\rm{total}}                  & = & \hat{k}_{{\rm{}e}}\,\mu_{{\rm{}e}}+\hat{k}_{{\rm{}p}}\,\mu_{{\rm{}p}}+\hat{k}_{{\rm{He}}^{2+}}\,\mu_{{\rm{He}}^{2+}},\label{eqn_mass_total}
\end{eqnarray}
where 
\mbox{$\hat{k}_{{\rm{}e}}, \hat{k}_{{\rm{}p}}$} and \mbox{$\hat{k}_{{\rm{He}}^{2+}}$} 
give the divvies of electrons, protons and $\alpha$-particles inside the plasma background
\mbox{i.e.} \mbox{$j\in\{{\rm{electron}}~({\rm{}e}), {\rm{proton}}~({\rm{}p}), \alpha{\rm{-particle}}~({\rm{He}}^{2+})\}$}.

The total velocity is affected 
both by the electric field and by Coulomb interaction namely as
\begin{eqnarray}
 \dt{v_{\rm{total}}} & = & -\frac{e E}{\mu_{\rm{total}}} \cos[\Theta_{\rm{total}}]+ \dt{v_{\rm{C,total}}}. \label{eqn_acceleration_total}
\end{eqnarray}
Again if one considers the same plasma background as above,
one obtains for the Coulomb acceleration, which is the
last term of Equation~(\ref{eqn_acceleration_total}) the following
expression
\begin{eqnarray}
 \dt{v_{\rm{C,total}}} & := & \dt{v^*_{\rm{}e}}+\dt{v^*_{\rm{}p}}+\dt{v^*_{{\rm{He}}^{2+}}}, \label{eqn_coulomacceleration_total}
\end{eqnarray}
where each addend is given according to Jackson~(1962) by
\begin{eqnarray}
 \dt{v^*_j} & = & - \frac{e^2 {N_j \rm{sign}}\left[v^*_j\right]}{4\pi\epsilon_0^2} \left(\frac{q_j}{\mu_{j}{v^*_{j}}}\right)^2 \ln\left[\sqrt{\frac{\lambda_{\rm D}^2+b_0^2}{2 b_0^2}}\right]. \label{eqn_coulombacceleration}
\end{eqnarray}
Hence, Equation~\ref{eqn_coulomacceleration_total} shows that the total Coulomb acceleration is
given by the sum of the electron, proton and \mbox{${\rm{He}}^{2+}$} accelerations.
Under the convention that 
\mbox{$v_j^*<0$} for an electron moving toward the Sun and 
\mbox{$v_j^*>0$} for one propagating outward, 
\mbox{${\rm{sign}}\left[v_j^*\right]$} in Equation~(\ref{eqn_coulombacceleration}) makes
sure, that the acceleration is in fact a deceleration of the electron
velocity. Therefore Equation~(\ref{eqn_coulombacceleration}) represents a kind of 
friction, which forces the electron to stop due to its energy loss per collision.
At last, the change of the radial component $r$ of the coordinate of the effective one-particle system's movement 
is given by
\begin{eqnarray}
 \dt{r}  & = & v_{{\rm{total}}} \cos\left[\Theta_{{\rm{total}}}\right] =: v_{{\rm{total}},\parallel}.\label{eqn_radialdistance}
\end{eqnarray}

If the 
initial altitude \mbox{$r_0:=r[t_0]$}, 
initial velocity \mbox{$v_{{\rm{total}},0}:=v_{{\rm{total}}}[t_0]$} and
initial pitch-angle \mbox{$\Theta_{{\rm{total}},0}:=\Theta_{{\rm{total}}}[t_0]$} 
with the introduced previous equations are given 
then it is possible 
to calculate the propagation of an electron
away from the acceleration site with the following 
final equations of motion:
\begin{eqnarray}
                                r[t+\dd{t}]  & = & v_{\rm{total}}[t] \cos\left[\Theta_{\rm{total}}[t]\right] \dd{t} \label{eqn_finaleqn_radial}\\
                 v_{{\rm{total}}}[t+\dd{t}]  & = & v_{\rm{total}}[t] + \left.\dt{v_{\rm{total}}}\right|_{t,\Theta_{\rm{total}}\left[t\right]} \dd{t} \label{eqn_finaleqn_velocity}\\
  \Theta_{{\rm{total}}}\left[t+\dd{t}\right] & = & \Theta_{\rm{total}}\left[t\right] + \left.\left(\dt{\Theta_{{\rm{total}}}} + {\mathcal{P}} \dt{\theta_{\rm{C,total}}}\right)\right|_{t,\Theta_{\rm{total}}\left[t\right]}\dd{t} \label{eqn_finaleqn_angle}
\end{eqnarray}
Here $\mathcal{P}$ denotes the result of a binary 
dice, whose results $-1$ and $1$ are equally distributed. This is needed to consider
the random Coulomb scattering directions. Equation~(\ref{eqn_finaleqn_angle}) explains
the relation between \mbox{$\Theta_{\rm{total}}$} and \mbox{$\theta_{\rm{C,total}}$}.
These presented Equations are sensitive to the models of the plasma background, i.e. 
the plasma densities, the magnetic fields and the electric fields. The geometry of 
the electric and the magnetic field has been arranged by assumption~\ref{assumption_collinear} 
on page~\pageref{assumption_collinear}.

Note that those equations allow a so-called backward calculation ($\dd{t}<0$).
Therefore this method provides a good possibility for estimating the altitude and 
the energies of the electrons both at the acceleration site (\mbox{$t=0=:t_0$}),
if for any \mbox{$t_{\rm{}f} > t_0$}
the altitude \mbox{$r[t_{\rm{}f}]$}, 
the velocity \mbox{$v_{{\rm{total}}}[t_{\rm{}f}]$} and 
the pitch-angle \mbox{$\Theta_{{\rm{total}}}[t_{\rm{}f}]$} are known.

These kind of information (as mentioned in the Section~\ref{sec_introduction}) can be obtained
\mbox{e.g.} by combining {\sl RHESSI} and {\sl WIND} data. Then, one has to assume 
a standard flare model, which implies that the acceleration site emits
the electrons in both directions (in- and outward) simultaneously.
\begin{figure}[t!]
   \begin{center}
    \includegraphics[width=0.49\textwidth]{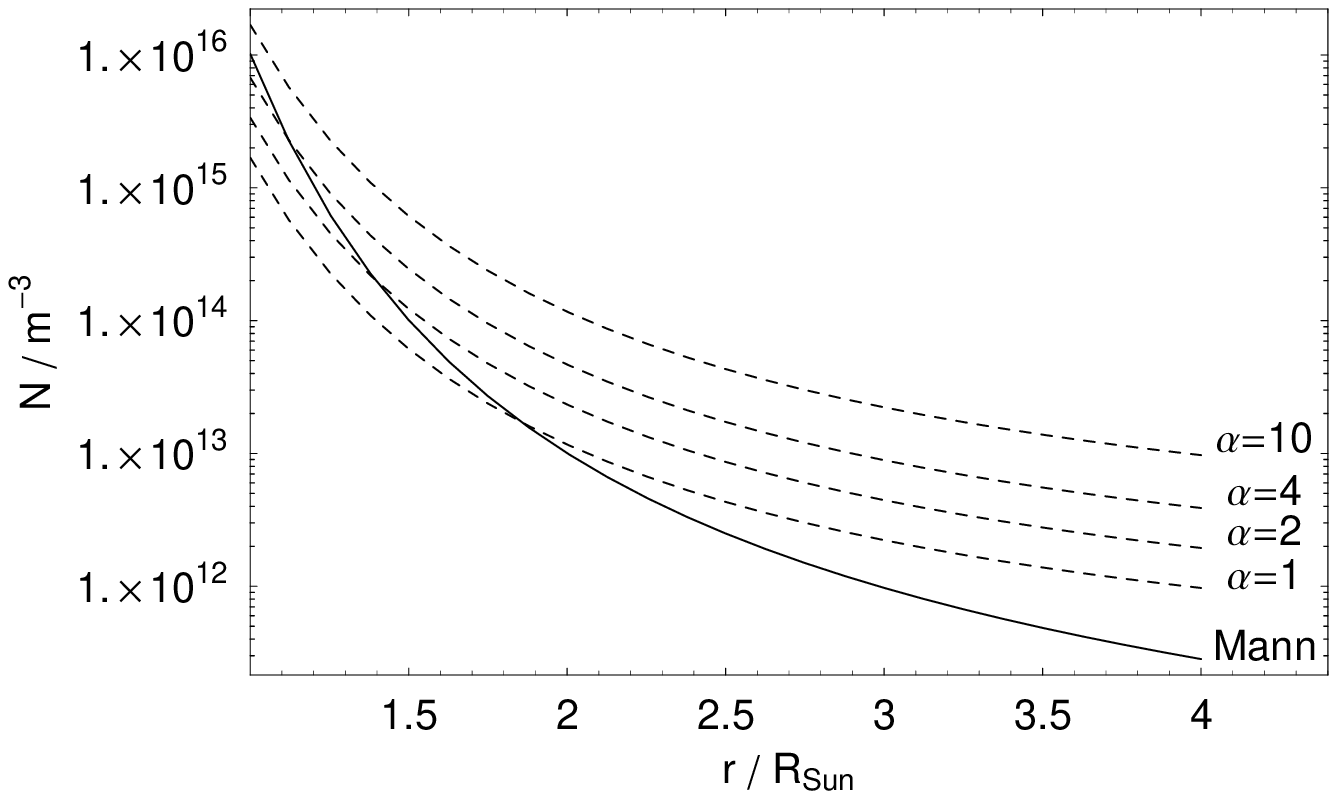} 
    \includegraphics[width=0.49\textwidth]{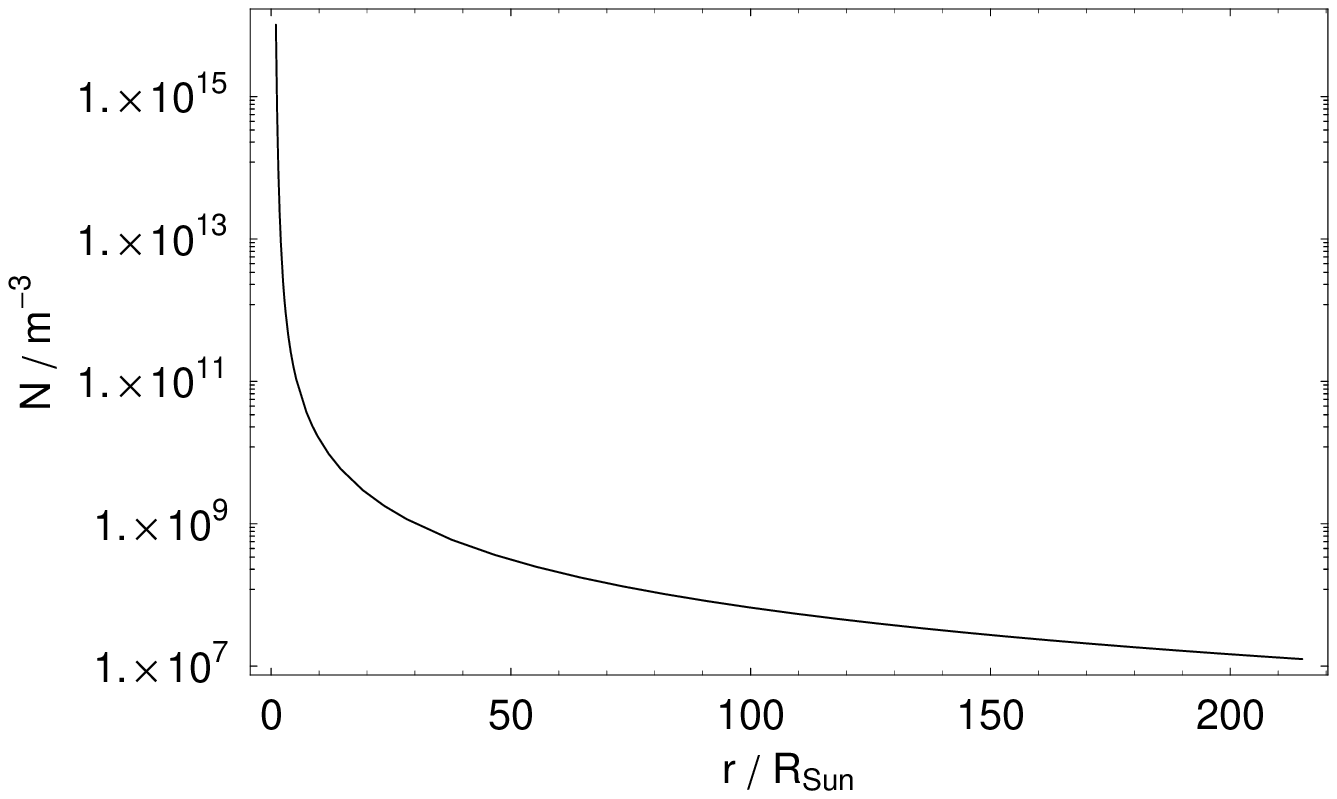}
    \caption[Atmospheric Model]{{Left:} The model of Newkirk~(1961), as given by Equation~(\ref{eqn_newkirk}), is plotted for several different $\alpha$s (dashed lines) 
                                        together with the introduced model by Mann~et~al.~(1999) (solid line). It is clearly visible that $r_{\rm S}$ is a function of $\alpha$.
                               {Right:} The density model according to Mann~et~al.~(1999) is plotted up to a range of \mbox{$1~{\rm AU}\approx{}215~\Rs$}.}
    \label{img_atmosphere}
   \end{center}
\end{figure}
\section{Plasma Background}\label{sec_background}
Within this Section the plasma background \mbox{i.e.}
the model for the densities, 
the global magnetic and electric fields
are introduced. 

\subsection{Density Models}
The composition of the background plasma is assumed to be fully ionised
and containing electrons, protons and $\alpha$-particles:
\mbox{$N_{\rm{}e}  =  \hat{k}_{{\rm{}e}} N \approx 0.52 N$}, 
\mbox{$N_{\rm{}p}  =  \hat{k}_{{\rm{}p}} N \approx 0.44 N$}, and
\mbox{$N_{{\rm He}^{2+}}  =  \hat{k}_{{\rm He}^{2+}} N \approx 0.04 N$},
where $N$ denotes the total particle number density,
which is the sum of the electron \mbox{$N_{\rm{}e}$}, 
the proton \mbox{$N_{\rm{}p}$} 
and the \mbox{${\rm He}^{2+}$} \mbox{$N_{{\rm He}^{2+}}$} densities.
Note that these quantities lead to a mean atomic weight 
of \mbox{$\tilde{\mu}=0.6$} as given in Priest~(1984).

For the coronal density model, one can use the $\alpha$-fold-Newkirk~(1961) model 
illustrated in Figure~\ref{img_atmosphere} and given with the following equation
\begin{eqnarray}
   N_{{\rm Newkirk}}[r,\alpha] & = & 4.2\alpha \hat{k}_{{\rm{}e}}^{-1} \times{}10^{10+4.32~\Rs/r} \quad{}{\rm m}^{-3},\label{eqn_newkirk}
\end{eqnarray}
while \mbox{$\Rs\approx 6.958 \times 10^8~{\rm m}$} denotes one solar radius. The
$\alpha$ parameter is supposed to be about $1$ in the case of the quite, 
about $4$ for the active and about $10$ for the super active Sun.
%
Due to the fact, that the Newkirk~(1961) model is only valid close to the
Sun, an interplanetary density model has to be taken into account additionally.
Mann~et~al.~(1999) introduced such a model, which considers the stationary flow of 
the solar wind, by solving the wind equation of Parker~(1981)
and the equation of continuity simultaneously
\begin{eqnarray}
  \frac{\hat{v}}{\hat{v}_{\rm{}c}^2} \ddfrac{\hat{v}}{r} & = & - \ddfrac{}{r}\big[\ln\left[N\right]\big] - 2 \frac{r_{\rm{}c}}{r^2}\label{eqn_windequation1}\\
                      N \hat{v} r^2 & = & 1.2\times 10^{34}~{\rm Hz},\label{eqn_windequation2}
\end{eqnarray}
with 
\mbox{$\hat{v}_{\rm{}c}:=(k_{\rm B} T \tilde\mu^{-1} m_{\rm{}p}^{-1})^{0.5}$} and 
\mbox{$r_{\rm{}c}:=0.5\,G M_{\mysun}\,\hat{v}_{\rm{}c}^{-2}$}.
Here $G$ stands for the Newtonian constant of gravitation, 
     \mbox{$M_{\mysun} \approx 1.985 \times 10^{30}~{\rm kg}$} represents the mass of the Sun 
     and $m_{\rm{}p}$ the mass of a proton, while $\hat{v}$ is the velocity of the solar wind.
     For a temperature of \mbox{$T=10^6~{\rm K}$} one obtains \mbox{$\hat{v}_{\rm{}c} \approx 120~{\rm Mm\,s^{-1}}$} and 
     \mbox{$r_{\rm{}c}\approx 4.2~\Rs$} while the numerical solution of the 
     Equations~(\ref{eqn_windequation1})~and~(\ref{eqn_windequation2}) 
     provide the interplanetary density model of Mann~et~al.~(1999), which is shown in 
     Figure~\ref{img_atmosphere}. 

To combine the presented models of Newkirk~(1961) and Mann~et~al.~(1999) one has
to introduce $r_{\rm S}$ as the point of intersection between both 
of them, then 
\begin{eqnarray}
 N[r,\alpha] = \left\{ {N_{{\rm Newkirk}}[r,\alpha] \atop {N_{{\rm Mann}}[r]}}\right.& {\rm for} &{{1~\Rs < r<r_{{\rm S}}}\atop{5~{\rm AU}> r\geq{}r_{{\rm S}}}},  \label{eqn_atmosphere}
\end{eqnarray}
with \mbox{$\alpha=4$} (and 
\mbox{$r_{\rm S}[\alpha=4]\approx 1.12~\Rs$}) 
yields to the density model, which is used within this paper (see \"O{}nel~(2004,2005) for more
information about this explained density model).

\subsection{Global Magnetic Field}\label{subsec_globalmagneticfield}
In the present paper the coronal magnetic field model $B_{{\rm D\&M}}$ of 
Dulk~\&~McLean~(1978) 
and the interplanetary magnetic field model $B_{{\rm M\&N}}$ of Marini~\&~Neubauer~(1990) 
by considering Musmann~et~al.~(1977) as well as Parker~(1958) are combined 
as elucidated in \"O{}nel~(2004).

This yields to a global magnetic field model 
which is plotted in Figure~\ref{img_bfield_efield} and 
described by the following equations:
\begin{eqnarray}
 B_{{\rm D\&M}}[r] & = & \left(\frac{\sqrt[3]{4} \, r}{\Rs}-\sqrt[3]{4}\right)^{-3/2} \times 10^{-4}~{\rm T}  \quad {\rm for~} 1.02~\Rs\la{}r\la 10~\Rs   \label{eqn_dulkmclean}\\
 B_{{\rm M\&N}}[r] & = & \sqrt{\frac{7.988~{\rm T}^2\,{\rm m}^4}{r^4} + \frac{9.6112\times{}10^7~{\rm T}^2\,{\rm m}^{2.2}}{r^{2.2}}} \quad {\rm for~} 10~\Rs\la{} r \la 1{\rm~AU} \label{eqn_marinineubauer} \\
 B[r] & = & \left\{ {B_{{\rm D\&M}}[r]} \atop {B_{{\rm M\&N}}[r]}\right. \quad {\rm for~} {{1.02~\Rs\la r\leq 10~\Rs}\atop{10~\Rs \leq r \la 1{\rm~AU}}} \label{eqn_magneticfluxdensitymodel}  
\end{eqnarray}

\subsection{Global Electric Field}\label{subsec_globalelectricfield}
\begin{figure}[t!]
   \begin{center}
    \includegraphics[width=0.49\textwidth]{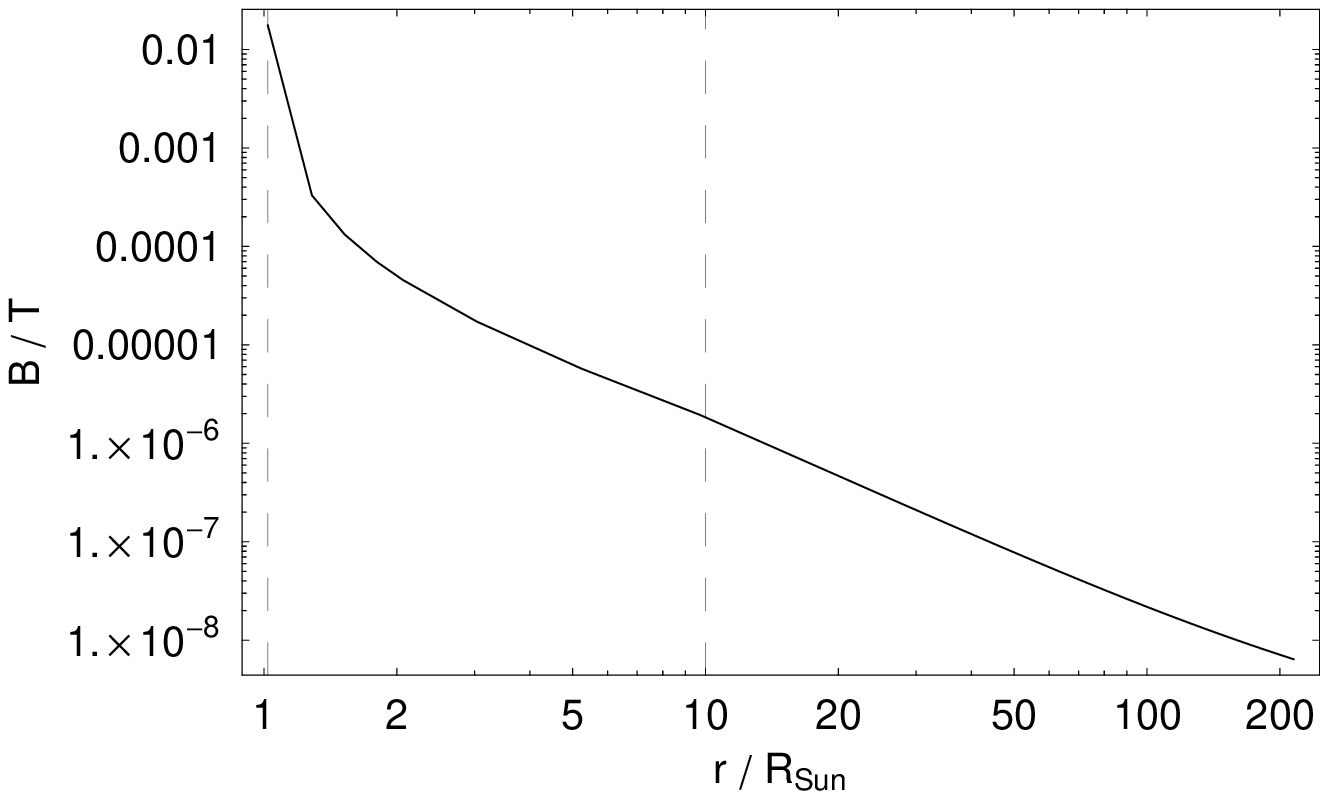} 
    \includegraphics[width=0.49\textwidth]{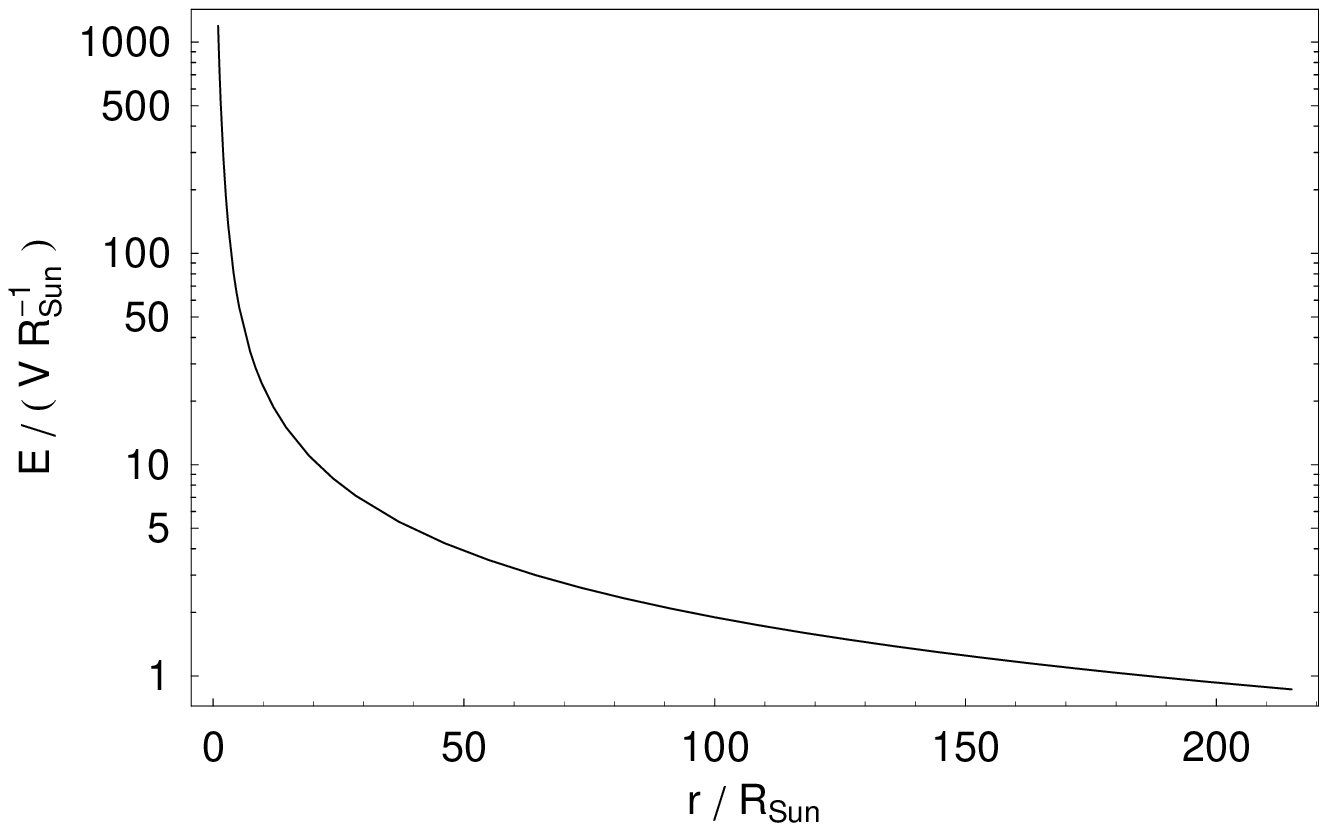}
    \caption[Electric and Magnetic Fields]{{Left:} The global magnetic field given by Equation~(\ref{eqn_magneticfluxdensitymodel}) is shown.
                                          {Right:} The global electric field given by Equation~(\ref{eqn_electricmodel}) is illustrated.}
    \label{img_bfield_efield}
   \end{center}
\end{figure}

If one considers the small mass of an electron compared
to every other ion, then, one can derive the way demonstrated in
\"O{}nel~(2004), a global electric field $E$, shown in Figure~\ref{img_bfield_efield},
from the spherical, hydrodynamical Euler 
equation for an ideal, isothermal electron fluid, if only the 
electrostatic force density is taken into account:
\begin{eqnarray}
 E[r] & = & -\frac{k_{\rm B} T}{e}\cdot\ddfrac{}{r}\Bigg[\ln\Big[N\big[r\big]\Big]\Bigg]\label{eqn_electricmodel}
\end{eqnarray}
Using Equation~(\ref{eqn_electricmodel}) one can calculate an electrostatic potential 
difference of \mbox{$1.8~{\rm kV}$} between the photosphere (\mbox{$r=1~\Rs$}) and the 
Earth (\mbox{$r = 1~{\rm AU}$}) by assuming the introduced density model Equation~(\ref{eqn_atmosphere}).
\section{Discussions} \label{sec_discussion}
\begin{figure}[t!]
   \begin{center}
    \hspace*{-0.07\textwidth}\includegraphics[width=1.12\textwidth]{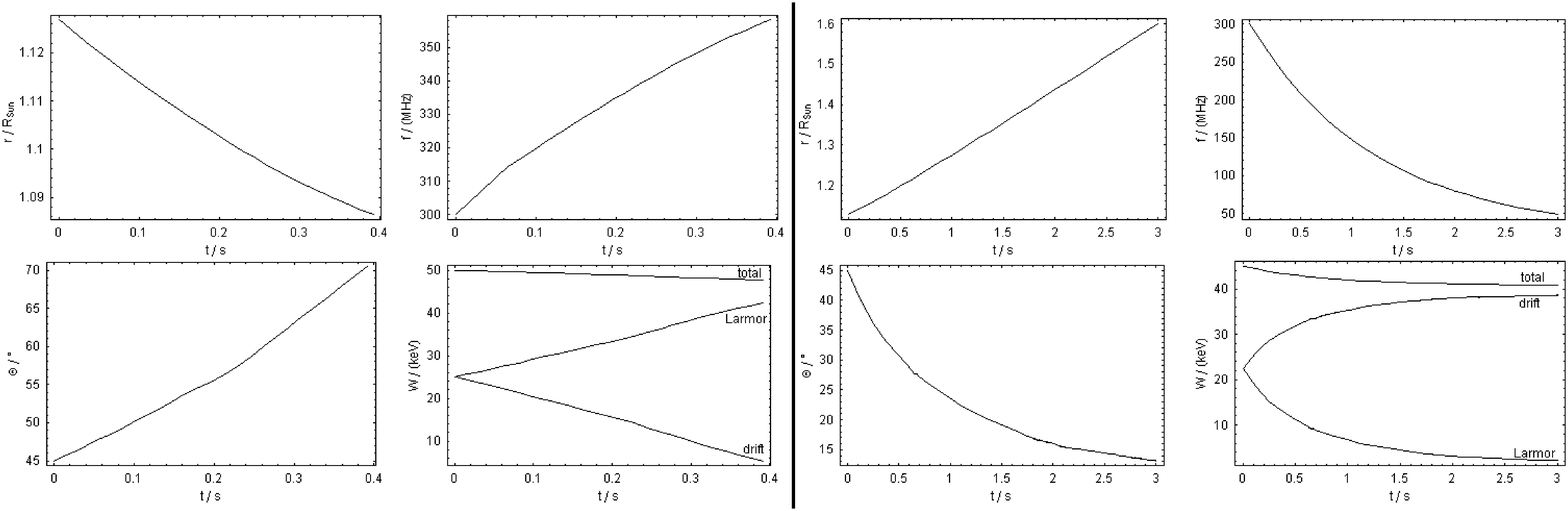} 
    \caption[Electron Transport]{Two examples for the electron transport model introduced in Section~\ref{sec_model} are shown. The diagrams are explained and discussed in Section~\ref{sec_discussion}.}
    \label{img_simulation_inandout}
   \end{center}
\end{figure}

Within this Section two diagrams (Figure~\ref{img_simulation_inandout}), 
one for the inner and one for the outer transport,
will be explained. These diagrams are not related to any particular solar event.
Their purpose is to show how those quantities
are effected by global magnetic and electric fields, and local 
Coulomb collisions.
Both diagrams in Figure~\ref{img_simulation_inandout} contain 
                       an altitude--time \mbox{$r[t]$},
                       electron plasma frequency (from Equation~(\ref{eqn_electronplasmafrequency}))--time \mbox{$f[t]$}, 
                       energy--time \mbox{$W[t]$} and a
                       pitch-angle--time \mbox{$\Theta[t]:=\Theta_{\rm{total}}[t]$}                     
evolution for an electron.
(Please keep in mind, that due to the chosen centre of mass system
those quantities describe only the effective one-particle 
system electron-plasma~background particle.)
In addition, the energy amount belonging to the velocity component parallel 
(drift energy) and perpendicular (Larmor energy) to the magnetic flux density 
is plotted within the energy--time diagram.
              
In both diagrams, 
based on usual our observations of type~III burst origins 
(\mbox{e.g.} Figure~\ref{img_aim}),
an electron at an initial altitude (of approximately \mbox{$r_0 \approx 1.127~\Rs{}$}), 
which corresponds to an electron plasma frequency of \mbox{$300~{\rm{MHz}}$},
with a total initial energy amount of \mbox{$50~{\rm{k}}e{\rm{V}}$},
and an initial pitch-angle of \mbox{$\Theta_{{\rm{total}},0}=0.25~\pi$} are assumed.
This means, that the initial velocity components 
parallel and perpendicular to the magnetic flux density are equal, 
so that at the acceleration site (\mbox{$t=t_0$}) the drift and the Larmor energies 
for the electron are exactly 
half of the total energy amount.

As it can be seen from Figure~\ref{img_simulation_inandout} 
the electron moving 
toward the Sun transfers its drift energy into Larmor energy due to the
magnetic (mirror) forces, while it loses energy due to Coulomb collisions.
The energy gain caused by the electric forces is insufficient to accelerate
the electron, which stops in its radial movement about \mbox{$0.4~{\rm{s}}$} later.
The electron moving toward the interplanetary
space behaves reversed: It transfers its Larmor energy into drift energy due to the
magnetic mirror forces, while it also loses energy due to Coulomb collisions and
the electric forces. Due to the decreasing particle density of the atmosphere and its
high drift energy, this electron is able to leave the solar atmosphere, with 
less energy loss.
Note, that the transport calculations are done $1\,000$~times for each transport
direction. Afterwards all single results for each direction have been averaged 
arithmetically in the shown Figure~\ref{img_simulation_inandout} to obtain a smooth plot.
\section{Summary} \label{sec_summary}
A 1-dimensional model for the propagation of energetic electrons
through a certain plasma background, with known average properties
has been introduced.

In case energetic electrons produced \mbox{e.g.} during solar flares 
can penetrate into the interplanetary space, they can be 
measured in-situ, \mbox{e.g.} by the {\sl WIND} spacecraft.
In addition, energetic electrons are held responsible for the non-thermal
X-ray radiation from the chromosphere, which can be observed by {\sl RHESSI} spacecraft.

Within this paper we performed a numerical study to characterise how 
the typical quantities during electron propagation could be affected, by 
considering average models for the local plasma density, the global 
electric and the magnetic field.

In our opinion, the main improvement of the deduced
model, which easily can be modified to treat proton or ion propagation
phenomena, compared with earlier common models
of the community (e.g. Brown 1972; Bai 1982) is the innovative
way of treating the pitch angle by introducing a binary dice.
\paragraph{{\it Acknowledgements.}}
   We appreciate the permission of the physics department 
   of the Technische Universit\"a{}t Berlin to use their
   computer facilities for our calculations.
   Furthermore the 
   authors would also like to express their
   thanks towards all members of the Solar Radio Group
   at the 
   AIP,
   and especially to R.~Miteva 
   and J.~Magdaleni\'c for the productive discussions. \newline
   In lovely memories of the physicist 
   Stephan Joachim Simon (TU--Berlin), who died on 
   \mbox{$05^{\rm{th}}\mbox{~April~}2005$} at the age
   of \mbox{$25$~\mbox{years}}.
\section*{References}
\everypar={\hangindent=1truecm \hangafter=1}

Bai,~T.,	
Transport of energetic electrons in a fully ionized hydrogen plasma,
{\it ApJ}, {\bf 259}, \mbox{341--349}, 1982.

Brown,~J.~C.,
The Directivity and Polarisation of Thick Target X-ray Bremsstrahlung from Solar Flares,
{\it Solar Phys.} {\bf 26}, \mbox{441--459}, 1972.

Dulk,~G.~A., and D.~J.~McLean,
Coronal magnetic fields,
{\it Solar Phys.}, {\bf 57}, \mbox{279--295}, 1978.

Jackson,~J.~D., 
{\sl Classical Electrodynamics}, 
John Wiley \& Sons, New York and London, \mbox{625--660}, 1962.

Newkirk,~G.~A.,  
The Solar Corona in Active Regions and the Thermal Origin of the Slowly Varying Component of Solar Radio Radiation.,
{\it ApJ}, {\bf 133}, \mbox{983--1013}, 1961.

Mann,~G., F.~Jansen, R.~J.~MacDowall, M.~L.~Kaiser, and R.~G.~Stone, 
A heliospheric density model and type III radio bursts, 
{\it A\&{}A}, {\bf 348}, \mbox{614--620}, 1999.

Mariani,~F., and F.~M.~Neubauer, 
The Interplanetary Magnetic Field,
in
{\sl Physics of the Inner Heliosphere I: Large Scale Phenomena}, 
edited by R.~Schwenn, and E.~Marsch, 
Springer-Verlag, Berlin, Heidelberg, \mbox{183--204}, 1990.

Melrose,~D.~B.,
Plasma emission mechanisms,
in
{\sl Solar Radiophysics: Studies of Emission from the Sun at Metre Wavelengths}, 
edited by D.~J.~McLean and N.~R.~Labrum, 
Camebridge University Press, Camebridge, \mbox{177--210}, 1985.

Musmann,~G., F.~M.~Neubauer, and E.~Lammers,
Radial variation of the interplanetary magnetic field between \mbox{$0.3~{\rm{AU}}$} and \mbox{$1.0~{\rm{AU}}$},
{\it JGZG}, {\bf 42}, \mbox{591--598}, 1977.

\"O{}nel,~H.,
{\sl Einfluss von Coulomb-St\"o{}\ss{}en auf die Ausbreitung von Elektronen im Flare-Plasma der Sonnenkorona},
Diploma Thesis at the Technische Universit\"a{}t Berlin, Center for Astronomy and Astrophysics in collaboration with the Astrophysikalisches Institut Potsdam,
Berlin, Germany, 2004.

\"O{}nel,~H., G.~Mann, and E.~Sedlmayr,
Propagation of Energetic Electrons in the Solar Corona and the Interplanetary Space,
in
{\sl Proceedings of the 11th European Solar Physics Meeting The Dynamic Sun: 
Challenges For Theory And Observations, 11-16 September 2005 Leuven, Belgium}, 
edited by D.~Dansey, S.~Poedts, A.~DE~Groof, and J.~Andries,
Published by ESA Publications, ESA/ESTEC, Noordwijk, NL, ISBN 92-9092-911-1, ISSN 1609-042X, 
ESA SP-600 CD-ROM, 2005.

Parker,~E.~N.,
Dynamics of the Interplanetary Gas and Magnetic Fields,
{\it ApJ}, {\bf 128}, \mbox{664--676}, 1958.

Parker,~E.~N.,
Photospheric flow and stellar winds,
{\it ApJ}, {\bf 251}, \mbox{266--270}, 1981.

Priest,~E.~R., 
{\sl Solar Magnetohydrodynamics}, 
D.~Reidel Publishing Company, Dordrecht, \mbox{82--83}, 1982.
\typeout{============================================================================}
\typeout{FINISHED: Contribution of H. Onel (onel@physik.tu-berlin.de - honel@aip.de) }
\typeout{============================================================================}
\end{document}